\newcommand{\la}{\langle}
\newcommand{\ra}{\rangle}
\newcommand{\lla}{\la\!\la}
\newcommand{\rra}{\ra\!\ra}

\newcommand{\beq}{\begin{eqnarray}}
\newcommand{\eeq}{\end{eqnarray}}
\renewcommand{\theequation}{\thesection.\arabic{equation}}

\newcommand{\do}{^{\circ }}
\newcommand{\cl}{\centerline}

\newcommand{\btem}{\bibitem}
\renewcommand{\TH}{T.\ Hatsuda}
\newcommand{\TK}{T.\ Kunihiro}
\newcommand{\PL}{Phys.\ Lett.\ {\bf B}}
\newcommand{\PTP}{Prog.\ Theor.\ Phys.}
\newcommand{\PR}{Phys.\ Rev.}
\newcommand{\PRL}{Phys.\ Rev. \ Lett.}

\newcommand{\HK}{T. Hatsuda and T. Kunihiro}
\newcommand{\KH}{T. Kunihiro and T. Hatsuda}

\documentstyle[12pt]{article}

\catcode`\@=11
\long\def\@makefntext#1{
\protect\noindent \hbox to 3.2pt {\hskip-.9pt
$^{{\ninerm\@thefnmark}}$\hfil}#1\hfill}		%CAN BE USED

\def\@makefnmark{\hbox to 0pt{$^{\@thefnmark}$\hss}}  %ORIGINAL

\def\ps@myheadings{\let\@mkboth\@gobbletwo
\def\@oddhead{\hbox{}
\rightmark\hfil\ninerm\thepage}
\def\@oddfoot{}\def\@evenhead{\ninerm\thepage\hfil
\leftmark\hbox{}}\def\@evenfoot{}
\def\sectionmark##1{}\def\subsectionmark##1{}}

\renewcommand{\thefootnote}{\fnsymbol{footnote}}

\newcounter{sectionc}\newcounter{subsectionc}\newcounter{subsubsectionc}
\renewcommand{\section}[1] {\vspace*{0.6cm}\addtocounter{sectionc}{1}
\setcounter{subsectionc}{0}\setcounter{subsubsectionc}{0}\noindent
	{\normalsize\bf\thesectionc. #1}\par\vspace*{0.4cm}}
\renewcommand{\subsection}[1] {\vspace*{0.6cm}\addtocounter{subsectionc}{1}
	\setcounter{subsubsectionc}{0}\noindent
	{\normalsize\it\thesectionc.\thesubsectionc. #1}\par\vspace*{0.4cm}}
\renewcommand{\subsubsection}[1]
{\vspace*{0.6cm}\addtocounter{subsubsectionc}{1}
	\noindent {\normalsize\rm\thesectionc.\thesubsectionc.
          \thesubsubsectionc.
	#1}\par\vspace*{0.4cm}}

\newcounter{appendixc}
\newcounter{subappendixc}[appendixc]
\newcounter{subsubappendixc}[subappendixc]

\renewcommand{\appendix}[1] {\vspace*{0.6cm}
        \refstepcounter{appendixc}
        \setcounter{figure}{0}
        \setcounter{table}{0}
        \setcounter{equation}{0}
        \renewcommand{\thefigure}{\Alph{appendixc}.\arabic{figure}}
        \renewcommand{\thetable}{\Alph{appendixc}.\arabic{table}}
        \renewcommand{\theappendixc}{\Alph{appendixc}}
        \renewcommand{\theequation}{\Alph{appendixc}.\arabic{equation}}
        \noindent{\bf Appendix \theappendixc #1}\par\vspace*{0.4cm}}

\def\abstracts#1{{
       \centering{\begin{minipage}{12.2truecm}
        \footnotesize\baselineskip=12pt\noindent
	\centerline{\footnotesize ABSTRACT}\vspace*{0.3cm}
	\parindent=0pt #1
	\end{minipage}}\par}}

\renewenvironment{thebibliography}[1]
	{\begin{list}{\arabic{enumi}.}
	{\usecounter{enumi}\setlength{\parsep}{0pt}
\setlength{\leftmargin 1.25cm}{\rightmargin 0pt}
	 \setlength{\itemsep}{0pt} \settowidth
	{\labelwidth}{#1.}\sloppy}}{\end{list}}

\newcounter{itemlistc}
\newcounter{romanlistc}
\newcounter{alphlistc}
\newcounter{arabiclistc}

\newcommand{\fcaption}[1]{
        \refstepcounter{figure}
        \setbox\@tempboxa = \hbox{\footnotesize Fig.~\thefigure. #1}
        \ifdim \wd\@tempboxa > 6in
           {\begin{center}
        \parbox{6in}{\footnotesize\baselineskip=12pt Fig.~\thefigure. #1}
            \end{center}}
        \else
             {\begin{center}
             {\footnotesize Fig.~\thefigure. #1}
              \end{center}}
        \fi}

\newcommand{\tcaption}[1]{
        \refstepcounter{table}
        \setbox\@tempboxa = \hbox{\footnotesize Table~\thetable. #1}
        \ifdim \wd\@tempboxa > 6in
           {\begin{center}
        \parbox{6in}{\footnotesize\baselineskip=12pt Table~\thetable. #1}
            \end{center}}
        \else
             {\begin{center}
             {\footnotesize Table~\thetable. #1}
              \end{center}}
        \fi}

\def\@citex[#1]#2{\if@filesw\immediate\write\@auxout
	{\string\citation{#2}}\fi
\def\@citea{}\@cite{\@for\@citeb:=#2\do
	{\@citea\def\@citea{,}\@ifundefined
	{b@\@citeb}{{\bf ?}\@warning
	{Citation `\@citeb' on page \thepage \space undefined}}
	{\csname b@\@citeb\endcsname}}}{#1}}

\newif\if@cghi
\def\cite{\@cghitrue\@ifnextchar [{\@tempswatrue
	\@citex}{\@tempswafalse\@citex[]}}
\def\citelow{\@cghifalse\@ifnextchar [{\@tempswatrue
	\@citex}{\@tempswafalse\@citex[]}}
\def\@cite#1#2{{$\null^{#1}$\if@tempswa\typeout
	{IJCGA warning: optional citation argument
	ignored: `#2'} \fi}}

 1
 1
 1

\font\ninerm=cmr9

\begin{document}

\centerline{\normalsize\bf CHIRAL SYMMETRY  AND $U_A(1)$ ANOMALY}
\baselineskip=22pt
\centerline{\normalsize\bf IN AN EFFECTIVE THEORY OF QCD\footnote{Invited
talk presented at Int. RCNP Workshop on Color Confinement and
          Hadrons, March 22 - 24, RCNP, Osaka, Japan. To be published in
           the proceedings from World Scientific.}}
%\baselineskip=16pt
%\centerline{\normalsize\bf MANUSCRIPT BY COMPUTER}

%\vfill
%\vspace*{0.6cm}
\centerline{\footnotesize TEIJI KUNIHIRO}
\baselineskip=13pt
\centerline{\footnotesize\it Faculty of Science and Technology,
 Ryukoku University}
\baselineskip=12pt
\centerline{\footnotesize\it Seta, Ohtsu-city, 520-21, Japan}
\centerline{\footnotesize E-mail: kunihiro@rins.ryukoku.ac.jp}

%\vfill
\vspace*{0.9cm}
\abstracts{We show on the basis of an effective theory of QCD
 that a wide variety
 of observables in the hadron world is governed by the chiral symmetry
together with an interplay between the axial anomaly and the explicit
symmetry
 breaking due to the current quark mass. We also discuss the nature of
 the chiral transition at finite temperature and related dynamical
phenomena  using the effective Lagrangian. Some phenomenological
 implications of the small vector coupling (``vector limit'')
at high temperatures are
 suggested.}
%\vspace*{0.6cm}
\normalsize\baselineskip=15pt
\setcounter{footnote}{0}
\renewcommand{\thefootnote}{\alph{footnote}}
\section{Introduction}
The main topic of this conference is the confinement of
 colored quarks and gluons.
 However, one may notice that even apart from the confinement,
 the basic properties of the hadron world include
(i) chiral symmetry and its dynamical and (small) explicit breaking,
(ii) $U_A(1)$ anomaly, (iii) approximate $SU_f(3)$ symmetry, (iv)
 the OZI rule and its violation and so on.
Some of these are interrelated.  The purpose of the present talk is
not to pin down the basic mechanism for the confinement, (i) and (ii)
 and so on  as other talks here do, but
to  show that an effective theory which embodies (i) and (ii) but not the
 confinement can well
describe a wide variety of phenomena and observables in the hadron world,
 and then
apply the effective theory to $T\not=0$ and/or $\rho\not=0$ systems.
 Thereby we shall elucidate the governing roles of (i) and (ii)
in the hadron
world and get insight into the nature of chiral transition
 at  $T\not=0$ and/or $\rho\not=0$.\footnote{A
good reference for this report may be found in the  review
 article\cite{HK94}.}
\ We shall also give heuristic
 discussions about phenomenological implications of the small coupling
 in the vector channel at high temperatures.

\section{The Model Lagrangian}
 Our model Lagrangian is the  generalized Nambu-Jona-Lasinio (NJL) model
with the anomaly term\cite{HK3a};
\beq
{\cal L}& =& \bar{q}i\gamma \cdot \partial q
+ \sum^{8}_{a=0}
{g_{_S} \over 2}[(\bar{q}\lambda_a q)^2 + (\bar{q}i\lambda_a
\gamma_5q)^2] -\bar {q}{\bf m}q+  g_{_D} [{\rm det}\bar{q}_i (1-\gamma_5)
q_j +h.c.],\nonumber \\
 \ \ \ &\equiv& {\cal L}_0  + {\cal L}_{SB} +{\cal L}_S + {\cal L}_{_{D}}\ ,
\eeq
where the quark field $q_i$ has three colors ($N_c=3$) and three flavors
($N_f=3$), $\lambda^a$ ($a$=0$\sim$8) are the Gell-Mann matrices with
$\lambda_0$=$\sqrt{2 \over 3}\bf{1}$. Here  ${\cal L}_0 +{\cal L}_S \equiv
{\cal L}_{NJL}$
 is the $U(3)$ generalization of the NJL model and has
 manifest flavor-$U_{_L}(3) \otimes U_{_R}(3)$ invariance.
 ${\cal L}_{_{SB}}$ is the explicit $SU_V(3)$ breaking part with
 $m_i$ ($i$=$u,d,s$) being the current quark mass.
Finally,
${\cal L}_{_{D}}$ in Eq.(1) denotes the term which
has $SU_{_L}(3) \otimes SU_{_R}(3)$  invariance but breaks the  $U_A(1)$
symmetry;
this term  is a reflection of the axial anomaly in QCD.
While ${\cal L}_S$ does not cause the flavor mixing, the anomaly term
does;
with the dynamical breaking of chiral symmetry,
  it induces effective 4-fermion vertices such as
$ \la\bar{d}d\ra (\bar{u}u)(\bar{s}s)$ and
$-\la\bar{d}d\ra (\bar{u}i \gamma_5 u)(\bar{s}i \gamma_5 s)$, where
 the former (latter) gives rise to
 a flavor mixing in the scalar (pseudo-scalar) channels.
 The fact that ${\cal L}_{_{D}}$ represents the  $U_{_A}(1)$ anomaly can be
seen in the anomalous divergence of the flavor singlet axial current
$J_5^{\mu}  = \bar{u}\gamma^{\mu}\gamma_5u+\bar{d}\gamma^{\mu}\gamma_5d+
\bar{s}\gamma^{\mu}\gamma_5s$ as
\beq
\partial_{\mu}J^{\mu}_5  =  -4N_f g_{_D} {\rm Im}({\rm det}\Phi)
+ 2i\bar{q}m\gamma_5 q ,
\eeq
which is to  be compared with the usual anomaly equation written
in  terms of the
topological charge density of the gluon field,
$\partial_{\mu} J^{\mu}_5= 2N_f {g^2 \over {32\pi^2}}
 F_{\mu \nu}^a \tilde{F}^{\mu \nu}_a + 2i \bar{q}m \gamma_5 q.$
Thus one may say that the determinantal 6-fermion operator
$-2g_{_D} {\rm Im}({\rm det}\Phi)$ simulates the effect
 of the gluon operator
 ${g^2 \over {32\pi^2}} F_{\mu \nu}^a \tilde{F}^{\mu \nu}_a$
 in the quark sector.

\bigskip

\begin{table}[h]
\tcaption{ Comparison of the theoretical estimates and the
experimental/empirical values of the basic physical quantities.  *
 indicates the quantity used as input.}\label{tab:tab1}
\small
\cl{
\begin{tabular}{||c|c|c||} \hline  \hline
& Theory & Empirical values  \\ \hline \hline
$M_u$ ($M_s$) & 335 (527)  & 336 (540) MeV \\
$\la \bar{u}u\ra^{NP}$ & $-(245)^3$   & $-(225\pm 25)^3$ MeV$^3$ \\
$\la \bar{s}s\ra^{NP}$/$\la \bar{u}u\ra^{NP}$ & 0.78  &  $0.8\pm 0.1$ \\
$m_{\pi}$ ($m_{K}$) & 138$^*$ (496$^*$)  & 138 (496) MeV\\
$m_{\eta}$ ($m_{\eta'}$) & 487 (958$^*$) & 549 (958) MeV\\
$m_{\sigma}$ ($m_{\sigma'}$) & 668 (1348) & $\sim 700$ ($\sim 1400$) MeV \\
$\Gamma_{\sigma \rightarrow 2 \pi}$  & $\sim 900$  & $\sim$Re
$m_{\sigma}$\\ $f_{\pi}$ ($f_{K}$) & 93.0$^*$ (97.7) & 93 (113) MeV\\
$f_{\eta}$ ($f_{\eta'}$) & 94.3 (90.8) & 93$\pm$9 (83$\pm$7) MeV \\
$\theta_{\eta}$ ($\varphi_{\sigma}$) & $-$21$^{\circ}$ ($-$6.8$^{\circ}$)
& $\sim -20^{\circ}$ ($-$) \\
$G_{\pi q}$ ($G_{Kq}$) & 3.5 (3.6) & $\sim 3.5$ ($-$) \\
$G_{\pi N}$ ($G_{\sigma N}$) & 12.7 (7 $-$ 10) & 13.4 ($\sim$10.0) \\
$\Sigma_{\pi N}$  & 49 $\pm 7$    & $45 \pm 10$ MeV\\
\hline \hline
\end{tabular}
}
\end{table}

\bigskip

In Table 1, we summarize  basic physical quantities
calculated in this model in the mean-field plus ring
approximation\cite{HK3a,HK3b};
 the  corresponding empirical
values are also quoted for comparison.
The numerical values are obtained with the following parameter set;
$\Lambda = 631.4 {\rm MeV},\ \   g_{_S}\Lambda^2=3.67,\ \
g_{_D}\Lambda^5=-9.29,
\ \   m_s=135.7{\rm MeV},$
where we have used a three-momentum cutoff scheme.
Here we remark  that the mixing angle $\theta _{\eta}$
of the $\eta$ and $\eta '$
 mesons is in a nice agreement with the experimental value; the mixing angle
 is a good measure of the strength of the axial anomaly.
 We also notice that the $\pi$-N sigma term $\Sigma _{\pi N}$ agrees with
 the ``empirical'' value. To calculate the baryon quantities such as
$\Sigma _{\pi N}$, we have employed the chiral quark model where the NJL
 model Lagrangian is adopted as a chiral Lagrangian\cite{HK3b}:
The constituent quark mass
 is  identified as that generated by the dynamical chiral symmetry breaking.
 Then a successful constituent quark model
(actually, Isgur-Karl's\cite{IK}) was employed.
Comments on other quantities listed in the table are given in  \cite{HK94}.
 Table 1 tells us  that the $SU(3)$ NJL model  reproduces the
fundamental physical quantities  in the accuracy of O(10\%-15\%).

\section{Static Properties at Finite $T$ and $\rho$}
 So much for the vacuum and hadron properties at zero temperature.
Now let us go to  finite temperature case.
Lattice simulations\cite{columbia} show that the order and even the
  existence
 of the phase transition(s) are largely dependent on the number
 of the flavors especially when the physical current quark masses
 are used: For $m_u\sim m_d\sim 10 {\rm MeV} << 100 {\rm MeV}
\stackrel {<}{\sim}m_s$, the phase transition may be weak 1st order
 or 2nd order or not exist.

The gross feature of the $T$ dependence and the striking difference
 between the condensates of u (d) quark and the s quark\cite{latticeus}
 can
 be well described by the NJL model\cite{HK87a,NPB}. It is noteworthy
that at high
 temperatures, the flavor $SU(3)_f$-symmetry gets worse badly\cite{NPB},
which may reflect in the baryon and the vector meson spectra, because
 they are well described by the constituent quark models.

Some people may be interested in the nature of the chiral transition in the
chiral limit in this model.
 The numerical calculation shows\cite{chilim} that the phase transition is
 of 2nd order, although  the thermodynamical potential is asymmetric with
respect to the zero condensate due to the cubic term coming from the
determinantal term.

How about the phase transition at finite density?\  A numerical
calculation\cite{NPB} shows
 that at low temperatures lower than about 50 MeV, the phase transition is
 of strong first order in the density direction.  Note that  our model
 Lagrangian has no vector term like $g_{_V}(\bar{q}\gamma _{\mu}q)^2$.
 As a matter of fact, the strength and even the existence of the
 1st order transition are strongly dependent on the strength of the
 vector coupling $g_{_V}$ \cite{asakawa}; the vector term prevents a
  high-density state.
\newpage
\section{Dynamic Properties}
\subsection{hadrons at $T<T_c$}
One can calculate  meson properties at finite temperature by computing the
 response functions.  In our model\cite{HK87},
  the $\sigma $ meson mass  decreases as $T$ increases till
 $T_c$, the critical temperature\footnote{$T_c$ may be defined as the
temperature at which $m_{\pi}$ starts to go high or the temperature where the
 quark condensate takes a half of its zero-temperature value.}:
$m_{\pi}$ is found to be constant as long as $T<T_c$.
 The lattice simulations  on the meson masses\cite{screening} show that
 both $m_{\pi}$ and $m_{\sigma}$
 show similar behaviors with those given in the NJL model.  One should
 note however that the masses in the lattice simulation are so called the
screening mass $m_{sc}(T)$,
  the exponent describing the mesonic correlations in the spatial
   direction;
 namely for  the hadronic operator ${O}_H$
\beq
\int dx dy d\tau \lla {O}_H(\vec{x}, \tau) {O}_H(\vec{0},0) \rra
 \rightarrow \exp [- m_{sc}(T) \mid z \mid ] \ \ \ \
 (\mid z \mid \rightarrow \infty) .
\eeq
A detailed calculation has shown that the two kind masses
 have a similar temperature dependence both in  the Wigner and the
  Nambu-Goldstone phases even off the chiral limit\cite{Friman}; see also
 Appendix of \cite{NPB}.
 This implies that the lattice results on the screening masses may suggest
 that the dynamical masses also behave like those given in the NJL model.

\subsection{Hadronic excitations at $T>T_c$}
 It is remarkable that there seem exist colorless hadronic excitations
even in the high-$T$ phase\cite{HK85,detar85} contrary to the naive
picture of it.\footnote{As for  the gluon sector,
it was shown that   non-perturbative effects seem
  significant in the low energy regime at  temperatures
$T_c < T< (2\sim 3)T_c$\cite{gluon}.}

Hatsuda and the present author noted that  there should exist
  precursory soft modes in the high temperature phase prior to
the phase transition if the chiral transition is
 of second order or weak first order:  The soft modes are actually
fluctuations of the order parameter of the phase transition,
 $\la\la(\bar q q)^2\ra\ra$ and  hence
 $\la\la(\bar q i\gamma _5\tau q)^2\ra\ra$
due to the chiral symmetry.
   They demonstrated these  using  an effective theory of QCD.\cite{HK85}

 The subsequent measurements of the screening masses on the
 lattice QCD supported the existence of such hadronic modes in the high-$T$
  phase \cite{screening}: Vector-mesonic and baryonic modes were obtained
   as well as the scalar and the pseudo-scalar mesons.
 The following comments  are in order here:
(i) The screening masses of the pion and the sigma meson are both well
below  $2\pi T$, which indicates that the
interactions between q-$\bar {\rm q}$ in the pseudo-scalar and
 the scalar channels are still rather strong even in the high-$T$
 phase, as suggested in \cite{HK85}.
(ii)
  The screening masses of the  vector modes coincide
  with $2\pi T$ within the error bars soon after the chiral restoration,
which may simply indicate that the interactions between q-$\bar {\rm q}$'s
 in this channel are absent or greatly suppressed\cite{EI}.
(iii) The nucleons exist as a parity doublet\cite{parity}, which
 had been considered to be unrealistic because such a model admitting
 a parity-doubled nucleons leads to the {\em vanishing} $\pi $-N coupling
 in the NG phase\cite{ben}.

We have seen that there is a reason d'etre of the pionic and sigma-mesonic
 excitations near the transition point even in the high-$T$ phase; they are
 the fluctuations of the order parameter of the chiral restoration.

 Then how   hadronic excitations in the vector channel?
 The approach of the hidden local symmetry\cite{hidden} seems
 to claim the existence of the vector mesons are intimately related
 with the chiral symmetry and its spontaneous breaking.
 The point (ii) above  suggests that there exist no strong correlations
in the  vector channel in the high-$T$ phase.
 The present author showed \cite{qnum} that
 the great suppression seen in the screening mass is consistent
 with the $T$-dependence of the quark-number susceptibility
$\chi _q(T)$
 obtained  by  the lattice simulations\cite{qnum2}:
As $T$ is raised   $\chi _q$ increases
   very rapidly around the critical point of the chiral transition.
We shall give
more  discussions on the susceptibility.

\subsection{Quark-number Susceptibility}
The quark-number  susceptibility $\chi _q$
 is  the measure of the response of the quark
  number density to infinitesimal changes in the chemical potentials
$\mu _i (i=u,d)$\cite{qnum};
\beq
\chi _q(T,\mu)&=& \Bigl[{\partial} /{\partial \mu _u}+
{\partial}/ {\partial \mu _d}\Bigr]( \rho_u+\rho _d),\nonumber \\
 \ \ &=& \beta \int d{\bf x}\lla \bar{q}(0,{\bf x})\gamma_0q(0,{\bf x})
\bar{q}(0,{\bf 0})\gamma_0q(0,{\bf 0})\rra,
\eeq
where $\mu =(\mu _u,\mu _d)$, $\hat N_q $ is the quark-number operator,
$\rho _i$  the i-th quark-number density,
 $V$ the volume of the system and $\beta=1/T$. In the following, we treat
 the case where $\mu_u=\mu_d\equiv \mu_q$.

  It is important that the number susceptibility
 at  finite density  is
 directly related to the (iso-thermal) compressibility
$\kappa _{_{T}}$ as $\kappa_{_T}={\chi_q}/{\rho ^2}$.
Therefore if $\chi_q$ of a system is large, the system is easy to
 compress, which  may be a reflection of a weak (if exists) repulsion between
the constituents of the system.

 In the simple free-quark gas model,
one sees that $\chi_q^{(0)}(T)$ increases  as the constituent
 quark mass $M(T)$
decreases owing to the partial chiral restoration
 and reaches $N_fT^2$ at $M(T)=0$. The enhancement is,however,
 found to be
  merely about 1.6 with $M(T)$ as described in the NJL model
  \cite{NPB}.
  Thus there must be an additional mechanism to increase  $\chi _q$
 to realize the anomalous (relative) enhancement obtained in the lattice
  simulations.

We  note here that $\chi_q$ is the density-density correlation which is
 nothing but the
 0-0 component of the vector-vector  correlations or fluctuations.
It means that the quark-number susceptibility is intimately
related with the properties of the vector mesons and the fluctuations
 in the vector channel.

  Using the NJL model with the vector coupling $g_{_V}$, one can show that
at $\mu_q=0$
\beq
\chi _q={{\chi _q^{(0)}(T)}\over {1+2g_{_V}\chi _q^{(0)}(T)}}
\eeq
where
$\chi _q^{(0)}(T)$ is the susceptibility for the free-quark gas.

In general, $\chi _q$ is suppressed  with the vector coupling because
 the denominator is  larger than unity for positive $g_{_V}$.
  One would  find it reasonable for a system at finite $\mu _q$
   because it implies that the system has a small compressibility;
 recall that $\chi_q$ is proportional to  $\kappa _T$.
 One thus sees the anomalous enhancement of
 $\chi_q$ as given by the lattice simulations can  be accounted for by
 a possible change (decrease) of  the vector coupling,\ i.e.,
  the vanishing or abrupt decrease after the chiral transition\footnote{
Recently, Brown
 and Rho\cite{br} have
 indicated that the chiral restoration may imply the decrease of the vector
 coupling, thereby realize the ``vector limit'' of Georgi\cite{georgi},
 on the basis of the work by Harada and Yamawaki\cite{yama}
 on the renormalization of the
 Lagrangian of  the hidden local symmetry.}.

When $\mu_q\not=0$
 there arises a  coupling between $\chi_q$ and the scalar-density
  susceptibility $\chi_s$ owing to the non-vanishing ``vector-scalar
 susceptibility" $\chi_{_{VS}}$, which are  defined by
\beq
\chi_s=-\frac {d \la\la\bar q q\ra\ra}{dm}
            =\beta \int d{\bf x}\la\la\bar q(0,{\bf x})q
             (0,{\bf x})\bar {q }(0,{\bf 0})q(0,{\bf 0})\ra\ra,
\eeq
\beq
\chi_{_{VS}}=\frac {\partial \la\la\bar {q }q\ra\ra}
{\partial \mu_q}
            =\beta \int d{\bf x}\la\la
            \bar {q}(0,{\bf x})\gamma _0 q (0,{\bf x})
\bar {q }(0,{\bf 0})q (0,{\bf 0})\ra\ra.
\eeq
 Thus when $\mu_q\not=0$,
 the fluctuation of the order parameter reflects in $\chi_q$.
This should give
 an enormous effect on $\chi_q$ when the chiral transition is of 2nd order
 or weak first order where the fluctuation of the order parameter becomes
 huge near the critical point. It would be intriguing to see this in
lattice calculations.

 Phenomenologically, the rise of $\chi_q$ means that
 the system is easy to compress as noted before, which implies that there
 develops a large density fluctuation near and  above critical temperature,
 especially in a finite-density system.\footnote{When $\mu_q\not=0$,
 it can be shown that $\chi_q\sim 1/g_{_V}$ hence blows up
 at $T$ near $T_c$.}
  The large density fluctuation
 may be reflected in the distribution of,  say, pions produced in the
 relativistic heavy ion collisions, and may become a seed for the generation
of heavy elements in the early universe.

\section{Summary and future problems}
We have seen that (to a surprise) the NJL model with the determinantal
interaction well describes the low-energy hadron world. This may imply that
 the low-energy behavior of the hadron world is mainly determined by the
chiral
 symmetry, and the axial anomaly and the explicit breaking due to the current
 quark masses give some variations to it.
 We have also seen that the model is useful to explore the finite temperature
 and/or density systems.

As future problems, one should, of course, incorporate the effect of the
confinement  explicitly together with the chiral symmetry, and see how and
 why the effect of the confinement did not affect so much the hadron
 phenomenology at low energy once the color-singlet states are projected out
 as in the NJL model.  The monopole condensation
 is enthusiastically advocated in this conference as the mechanism of the
color confinement. The chiral symmetry breaking is probably related with the
other topological object, instanton. One may be interested in how the
monopole
 instanton configuration is affected with the presence of the dynamical
quarks,
 especially light quarks:  The proposed picture of the confinement based on
the
 monopole condensation seems to be  extracted when quarks are heavy.

In describing baryons, we have
employed a chiral quark model where the NJL model
is used as a chiral Lagrangian together with the perturbative gluon exchange.
 It would be interesting to apply the renormalization group to it and see
 if the Lagrangian becomes the perturbative QCD with the NJL coupling being
 vanishingly small at high-energy scales.

{\cl{\bf Acknowledgements}}
I would like to acknowledge Tetsuo Hatsuda for his collaboration on the works
 which some part of this report is based on.
This work was supported by the Japanese
 Grant-in-Aid for Science Research Fund of the Ministry of Education, Science
 and Culture, No. 05804014 and Joint Research Center for Science and
Technology, Ryukoku University.  This whole manuscript was written
 when I stayed in Institute for Nuclear Theory (INT) at the University of
Washington.
 I thank INT for its hospitality and the Department of energy for partial
 support.

\bigskip

{\cl{\bf References}}

\end{document}